\documentstyle[aps,prl,twocolumn,epsfig]{revtex}

\begin{document}

\twocolumn[\hsize\textwidth\columnwidth\hsize\csname
@twocolumnfalse\endcsname

\title{High current-carrying capability in c-axis-oriented superconducting 
MgB$_2$ thin films}
\author{Hyeong-Jin Kim, W. N. Kang$^*$, Eun-Mi Choi,
 Mun-Seog Kim, Kijoon H.P. Kim, and Sung-Ik Lee}
\address{National Creative Research Initiative Center for Superconductivity\\
and Department of Physics, Pohang University of Science and Technology,\\
Pohang 790-784, Republic of Korea }
\draft
\maketitle

\begin{abstract}
{In high-quality c-axis-oriented MgB$_{2}$ thin films, we observed high critical 
current densities $(J_{c})$ of $\sim $ 16 MA/cm$^{2}$ at 15 K under self fields 
comparable to, and exceeding, those of cuprate high-temperature superconductors. 
The extrapolated value of $J_{c}$ at 5 K was estimated to be 
$\sim $ 40 MA/cm$^{2}$.  At a magnetic field of 5 T, 
a $J_{c}$ of $\sim $ 0.1 MA/cm$^{2}$ was detected at 15 K, suggesting that 
this compound would be a very promising candidate for practical applications 
at high temperature and lower power consumption. The vortex-glass phase
 is considered to be a possible explanation for the observed high current carrying
 capability.}
\end{abstract}

\vskip 2.5pc]

The recent discovery of the binary metallic MgB$_{2}$ superconductor \cite
{Nagamatsu01} with a remarkably high transition temperature $T_{c}$ = 39 K
has attracted great interest in both basic scientific \cite
{Budko01,Finnemore01,Karapetrov01,An01,Kang01} and practical applications 
\cite
{Labalestier01,Jung01,Paitnaik01,Soltanian01,Kang01-1,Canfield01,Bugoslavsky01,Eom01}%
. This new compound is expected to be useful for superconducting magnets and
microelectronic devices at low cost because its transition temperature is 2 $%
-$ 4 times higher than those of conventional metallic superconductors such
as Nb$_{3}$Sn and Nb-Ti alloy. The strongly linked nature of the intergrains 
\cite{Labalestier01} with a high charge carrier density\cite{Kang01} in this
material is a further indication of its possible use in technological
applications. Recently, an upper critical field, $H_{c2}(0),$ of 29 $\sim $
39 T \cite{Jung01,Paitnaik01}, which was much higher than previously
reported, was observed, suggesting that MgB$_{2}$ should be of considerable
use for practical application in superconducting solenoids using mechanical
cryocoolers, such as closed-cycle refrigerator. In addition to the higher $%
T_{c}$ and $H_{c2}$ in MgB$_{2}$, the magnitude of the critical current
density is a very important factor for practical applications. For example,
if a superconducting wire carries a high electric power, the size of the
cryogenic system can be reduced considerably so that the system can operate
with lower power consumption. Indeed, the successful fabrication of Fe-clad
MgB$_{2}$ tape has been reported \cite{Soltanian01}. This tape showed a $%
J_{c}$ of 1.6 $\times $ 10$^{4}$ A/cm$^{2}$ at 29.5 K under 1 T, which is
encouraging for practical application of MgB$_{2}$.

In order to explain the nature of the vortex state in strong magnetic field
for cuprate high-$T_{c}$ superconductors (HTS), Fisher $et$ $al.$\cite
{Fisher89} proposed the theory of vortex-glass superconductivity by
considering both the pinning and the collective effects of vortex lines.
According to this theory, a diverging vortex glass correlation length $(\xi
) $ near the vortex-glass transition $(T_{g})$ can be described by $\xi \sim
|T-T_{g}|^{-\nu }$ and a correlation time scale $\xi ^{{\rm z}}$, where $\nu 
$ is a static exponent and z is a dynamic exponent; thus, $I-V$ curves can
be expressed by universal scaling functions. For HTS, experimental evidence
of a vortex glass phase has been reported \cite{Koch89}. Moreover, a
vortex-glass transition was observed in an untwinned single crystal of YBa$%
_{2}$Cu$_{3}$O$_{7}$ after inducing a sufficiently high density of pinning
centers, suggesting that a vortex-glass phase may be one origin of the high $%
J_{c}$\cite{Petrean00}.

In this Letter, we report a high current-carrying capability in high-quality
MgB$_{2}$ thin films, which was confirmed by direct current-voltage $(I-V)$
measurements for various magnetic fields and temperatures. Furthermore, the
vortex glass phase will be discussed as a possible origin of the high $J_{c}$
in MgB$_{2}$ thin films.

The MgB$_{2}$ thin films were fabricated using a two-step method; the
detailed process is described elsewhere\cite{Kang01-1}. Briefly, an
amorphous B thin film was deposited on a (1 $\bar{1}$ 0 2) Al$_{2}$O$_{3}$
substrate at room temperature by using pulsed laser deposition. The B thin
film was put into a Nb tube together with high purity Mg metals (99.9\%) and
the Nb tube was then sealed using a arc furnace in an Ar atmosphere. The
heat treatment was carried out at 900 C for 10 - 30 minutes in an evacuated
quartz ampoule, which was sealed under high vacuum. The film thickness was
0.4 $\mu $m, which was confirmed by scanning electron microscopy. X-ray $%
\theta -2\theta $ diffraction patterns indicated that the MgB$_{2}$ thin
film had a highly c-axis-oriented crystal structure normal to the substrate
surface; no impurity phase was observed. The $\phi $-scan x-ray diffraction
patterns showed randomly oriented crystal structures along ab-plane of the
thin film. In order to measure the $I-V$ characteristics, we used standard
photolithography, and then chemical etching in an acid solution, HNO$_{3}$ $%
(50\%)$ and pure water $(50\%)$, to pattern the thin films into microbrige
shapes (inset of Fig.\ref{fig:fig1}) with strip dimensions of 1 mm long and
65 $\mu $m wide. To obtain good ohmic contacts $(<\ 1\ \Omega )$, we coated
the contact pads with Au films after using Ar ion-beam milling to clean the
film surface. This patterning process didn't degrade the superconducting
properties of the MgB$_{2}$ thin films.

Figure \ref{fig:fig1} shows the typical temperature dependence of the
resistivity of a MgB$_{2}$ thin film measured after patterning into a
microbridge shape. An onset transition temperture of 39 K with a very sharp
transition of $\sim $ 0.2 K, determined from the 90$\%$-to-10$\%$ drop off
of the normal-state resistivity, was observed. The observed room-temperature
(300 K) resistivity of 11.9 $\mu \Omega $ cm for the thin film was similar
to that in a polycrystalline MgB$_{2}$ wire\cite{Canfield01}, and a residual
resistivity ratio, RRR = $\rho $(300K)/$\rho $(40K), of 2.3, which is much
smaller than the value in the MgB$_{2}$ wire, was observed. This large
difference between the RRR values depends on the synthesis method, and its
cause is still under debate\cite{Kang01,Eom01}. A very small (less than 0.5$%
\%$) magnetoresistance was observed at 5 T and 40 K.

We used a superconducting quantum interference device magnetometer (SQUID,
Quantum Design) to measure the magnetization $(M$-$H)$ hysteresis loops of
MgB$_{2}$ thin films in the field range of -5 T $\leq H\leq $ 5 T with the
field parallel to the {\it c} axis. Figure \ref{fig:fig2} shows the $M$-$H$
curves at temperatures of 5, 15, and 35 K. Below $T$ = 10 K, the
magnetization at low field decreases with decreasing temperature (lower
panel of Fig. \ref{fig:fig2}), indicating a dendritic penetration of
vortices. This may be explained by a thermomagnetic instability in the flux
dynamics \cite{Johansen01}. Therefore, we may not apply the Bean critical
state model in this temperature region.

Figure \ref{fig:fig3} shows $J_{c}$, estimated from the $M$-$H$ loops (open
symbols) and measured directly by using a tranport method (solid symbols),
as a function of temperatures for various magnetic fields. The transport $%
J_{c}$ was determined by using a voltage criterion of 1 $\mu $V/mm. We
calculated the values of $J_{c}$ from the $M$-$H$ curves, by using the Bean
critical state model $(J_{c}=30\Delta M/r)$, where $\Delta M$ is the height
of $M$-$H$ loops. Here, we used $r$ = 1.784 mm, which is the radius
corresponding to the total area of the sample size, and was calculated from $%
\pi r^{2}$ = 4 $\times $ 2.5 mm$^{2}$. With this sample size, the $J_{c}$
curves obtained from the $M$-$H$ loops and the $I$-$V$ measurements
coincided, indicating the strongly linked nature of the intergrains on the
thin film; this behavior is different from that of the HTS\cite{Krusin95}.
Under a self field, the $J_{c}$ was $\sim $ 16 MA/cm$^{2}$ at 15 K. This
value is higher than the $J_{c}$ of 10 MA/cm$^{2}$ observed in
polycrystalline MgB$_{2}$ films grown on (0001) Al$_{2}$O$_{3}$ and (100)
MgO substrates\cite{Moon01}. As mentioned before, since the critical state
model cannot be applied to the temperature region below 15 K, the transport $%
J_{c}$ at 5 K is probably higher than that estimated by the Bean critical
state model. From $I-V$ measurements using polycrystalline thin film, a
monotonic increase of the critical current density with decreasing
temperature was observed at low temperature\cite{Moon01}. Based on the
temperature dependence of $J_{c}$ measured at 0.5 T, the extrapolated value
of $J_{c}$ at 5 K was estimated to be $\sim $40 MA/cm$^{2}$. This value is
comparable to that of YBa$_{2}$Cu$_{3}$O$_{7}$ thin film \cite{Roas90}, and
even exceeds the values for other HTS, such as Hg- and Bi-based
superconductors\cite{Kang98,Yamasaki93}. The high $J_{c}$ of $\sim $ 0.1
MA/cm$^{2}$ at 37 K under a self field suggests that MgB$_{2}$ thin films
have very high potential for low-cost applications in electronic devices
operating at high temperature, such as microwave devices and portable SQUIDs
sensors, by using miniature refrigerators. At $H$ = 5 T, the
current-carrying capability of 0.1 MA/cm$^{2}$ at 15 K may be of
considerable importance for practical applications in superconducting
solenoids using mechanical cryocoolers with low power consumption, if we can
fabricate high-quality MgB$_{2}$ thick films or tapes.

In order to investigate the vortex-phase diagram of MgB$_{2}$ thin films, we
measured the $I$-$V$ characteristics for various magnetic fields, as shown
in Fig. \ref{fig:fig4}. The $I$-$V$ curves in the upper inset of Fig. \ref
{fig:fig4} are very similar to those features of YBa$_{2}$Cu$_{3}$O$_{7}$
superconductor\cite{Koch89} around the vortex-glass transition temperature $%
T_{g}$. According to the vortex-glass theory\cite{Fisher89}, $I$-$V$ curves
show positive curvature for $T>T_{g}$, negative curvature for $T<T_{g}$, and
a power-law behavior at $T_{g}$, which is in good agreement with our results
with $T_{g}$ = 26.15 K at $H$ = 3 T. Furthermore, near $T_{g}$, these $I$-$V$
curves can be described by a universal scaling function with two common
variables, $V_{sc}=V/I|T-T_{g}|^{\nu ({\rm z}-1)}$ and $%
I_{sc}=I/T|T-T_{g}|^{2\nu }$. All the $I$-$V$ curves collapse onto a scaling
function with a static exponent of $\nu $ = 1.0 and a dynamic exponent of z
= 4.5. These values for the exponents are in good agreement with the
theoretical predictions for a three-dimensional (3D) system. This scaling
behavior is also followed by the $I$-$V$ curves measured at other fields
from 1 to 5 T. The bottom inset of Fig. \ref{fig:fig4} shows the phase
diagram in the $H$-$T$ plane. The $H_{c2}(T)$ were estimated from the $R$-$T$
curves when the resistivity drops to 90$\%$ of the normal-state resistivity.
We find that the vortex-glass region of MgB$_{2}$ is wide, implying that the
pinning force is very strong at low temperature. We suggest that the high
current-carrying capability of the MgB$_{2}$ superconductor probably
originates from a 3D vortex-glass phase with strong pinning disorder, and
from a higher density of charge carriers \cite{Kang01}. Indeed, the
vortex-glass phase of untwinned YBCO single crystals was observed only for
high disordered samples after proton irradiation whereas a vortex-lattice
melting transition was observed in pristine samples \cite{Petrean00}.

In summary, we have studied $J_{c}$ in MgB$_{2}$ thin films by using both
the $M$-$H$ hysteresis and $I$-$V$ measurements. We find that these two set
of data collapse quite well into one curve over the entire temperature
region, indicating the strongly linked current flow in this material. For a
magnetic field of 5 T, a critical current density of $\sim $ 0.1 MA/cm$^{2}$
was detected at 15 K, suggesting that this compound is a very promising
candidate for practical applications at high temperature, such as
liquid-He-free superconducting magnet systems and superconducting electronic
devices, and using mechanical or miniature cryocoolers with lower power
consumption. We suggest a 3D vortex-glass phase as a possible origin for the
high current-carrying capability of MgB$_{2}$.

\acknowledgements
This work is supported by the Creative Research Initiatives of the Korean
Ministry of Science and Technology.

\begin{figure}[tbp]
\caption{Resistivity vs. temperture for an MgB$_2$ thin film grown on an Al$%
_2$O$_3$ substrate by using pulsed laser deposition with post annealing. The
inset shows the narrow bar pattern, 65 $\protect\mu$m $\times$ 1 mm, of the
MgB$_2$ thin film.}
\label{fig:fig1}
\end{figure}

\begin{figure}[tbp]
\caption{Upper part shows the $M-H$ hysteresis loop at 5 K (solid circles),
15 K (open circles), and 35 K (triangles). The lower is a magnified view of
the low-field region at 5 and 15 K.}
\label{fig:fig2}
\end{figure}

\begin{figure}[tbp]
\caption{Temperature dependence of the critical current density of MgB$_2$
thin films for $H$ = 0 $-$ 5 T extracted from the $M-H$ (open symbols) and
the $I-V$ (solid symbols) curves. $J_c$ = 0.1 MA/cm$^2$ is a common
benchmark for practical applications.}
\label{fig:fig3}
\end{figure}

\begin{figure}[tbp]
\caption{Vortex-glass scaling behavior. When two variables, $%
V_{sc}=V/I|T-T_{g}|^{\protect\nu ({\rm z}-1)}$ and $I_{sc}=I/T|T-T_{g}|^{2%
\protect\nu }$, are used, the $I-V$ curves collapse into a scaling function
near the vortex-glass phase transition temperature. The upper inset shows
the $I-V$ characteristics for $T$ = 24.8 $-$ 28 K in 0.2 K steps under a
field of $H$ = 3 T. The lower inset shows the phase diagram based on a
vortex-glass (VG) to vortex-liquid (VL) transition.}
\label{fig:fig4}
\end{figure}

\end{document}